\documentclass[11pt]{article}

\usepackage[textwidth=15.5cm,textheight=22.5cm]{geometry}
\usepackage{bm}
\usepackage{amsmath,amssymb,color,graphicx,amscd,amsfonts}
\usepackage{latexsym}
\usepackage{graphicx}
\usepackage{cite}
\usepackage{subfig}
\usepackage{bbm}
\usepackage{ulem}

\def\diag{\mathop{\rm diag}\nolimits}
\def\sdet{\mathop{\rm sdet}\nolimits}
\def\Re{\mathop{\rm Re}\nolimits}
\def\Im{\mathop{\rm Im}\nolimits}

\def\da{\dot{a}}
\def\db{\dot{b}}
\def\dc{\dot{c}}
\def\dd{\dot{d}}

\def\bA{\bar{A}}
\def\bB{\bar{B}}
\def\bC{\bar{C}}
\def\bD{\bar{D}}

\def\FullSG{PSU(2,2|4)}

\def\SuSp{\frac{PSU(2,2|4)}{SO(4,1)\times SO(5)}}

\def\dI{\dot{I}}
\def\dJ{\dot{J}}

\def\a{\alpha}
\def\b{\beta}
\def\d{\delta}

\def\s{\sigma}

\allowdisplaybreaks[4]

\def\sla{\raise.15ex\hbox{$/$}\kern-.57em}

\allowdisplaybreaks[1]

\newcommand{\bdm}{\begin{displaymath}}
\newcommand{\edm}{\end{displaymath}}

\def\b{\beta}
\def\a{\alpha}

\def\s{\sigma}

\def\d{\delta}

\newcommand{\ie}{{\it i.e.\ }}

\DeclareMathAlphabet{\mathpzc}{OT1}{pzc}{m}{it}

\makeatletter
\newif\if@borderstar
\def\bordermatrix{\@ifnextchar*{%
 \@borderstartrue\@bordermatrix@i}{\@borderstarfalse\@bordermatrix@i*}%
}
\def\@bordermatrix@i*{\@ifnextchar[{\@bordermatrix@ii}{\@bordermatrix@ii[()]}}
\def\@bordermatrix@ii[#1]#2{%
\begingroup
 \m@th\@tempdima8.75\p@\setbox\z@\vbox{%
 \def\cr{\crcr\noalign{\kern 2\p@\global\let\cr\endline }}%
 \ialign {$##$\hfil\kern 2\p@\kern\@tempdima & \thinspace %
  \hfil $##$\hfil && \quad\hfil $##$\hfil\crcr\omit\strut %
  \hfil\crcr\noalign{\kern -\baselineskip}#2\crcr\omit %
  \strut\cr}}%
 \setbox\tw@\vbox{\unvcopy\z@\global\setbox\@ne\lastbox}%
 \setbox\tw@\hbox{\unhbox\@ne\unskip\global\setbox\@ne\lastbox}%
 \setbox\tw@\hbox{%
  $\kern\wd\@ne\kern -\@tempdima\left\@firstoftwo#1%
  \if@borderstar\kern 2pt\else\kern -\wd\@ne\fi%
 \global\setbox\@ne\vbox{\box\@ne\if@borderstar\else\kern 2\p@\fi}%
 \vcenter{\if@borderstar\else\kern -\ht\@ne\fi%
  \unvbox\z@\kern -\if@borderstar2\fi\baselineskip}%
 \if@borderstar\kern-2\@tempdima\kern2\p@\else\,\fi\right\@secondoftwo#1 $%
 }\null \;\vbox{\kern\ht\@ne\box\tw@}%
\endgroup
}
\makeatother
\date{}

\begin{document}

\thispagestyle{empty}

\setcounter{page}{0}

\begin{flushright} 
OIQP-15-6\\
%\today\\
%\currenttime\\

\end{flushright} 

\vspace{0.1cm}

\begin{center}
{\LARGE
Simple variables for AdS$_5 \times S^5$ superspace
\rule{0pt}{20pt}  }
\end{center}

\vspace*{0.2cm}

\renewcommand{\thefootnote}{\alph{footnote}}

\begin{center}

Hidehiko S{\sc himada}\,\footnote
         {
E-mail address: shimada.hidehiko@gmail.com} \\
\vspace{0.3cm}
{\it Okayama Institute for Quantum Physics, Okayama, Japan}\\

\end{center}

\vspace{1cm}

\begin{abstract}
We introduce simple variables
for describing the AdS$_5\times S^5$ superspace,
\ie $\frac{PSU(2,2|4)}{SO(4,1)\times SO(5)}$.
The idea is to embed the coset superspace
into a space described by variables which are in linear (ray) representations 
of the supergroup $PSU(2,2|4)$ by imposing certain supersymmetric quadratic constraints
(up to two overall U(1) factors).
The construction can be considered as a supersymmetric generalisation 
of the elementary realisations of 
the $AdS_5$ and the $S^5$  spaces
by the SO(4,2) and SO(6) invariant quadratic constraints on 
two six-dimensional flat spaces.
\end{abstract}

%%%%%%%%%%%%%%%%%%%%%%%%%%%%%%%%%%%%%%%
%%%%%%%%%%%%%%%%%%%%%%%%%%%%%%%%%%%%%%%
%%%%%%%%%%%%%%%%%%%%%%%%%%%%%%%%%%%%%%%

\newpage

\setcounter{footnote}{0}

\renewcommand{\thefootnote}{\arabic{footnote}}

String theory in AdS$_5\times S^5$ 
has been studied extensively in 
recent years,
because of the AdS/CFT correspondence~\cite{RBMaldacena}.
The theory is also a prime example of string theories with
non-zero Ramond-Ramond fields in their backgrounds,
and has a very high degree of symmetry,
in particular, the maximal supersymmetry $PSU(2,2|4)$.
The classical action of the theory~\cite{RBMetsaevTseytlin}
in the Green-Schwarz formalism~\cite{RBGreenSchwarz, 
RBGrisaruHoweMezincescuNilssonTownsend}
describes the propagation of the strings in the target superspace
$\SuSp$, which is expressed as a coset space.

It is the purpose of the present note
to point out that this superspace has a simple realisation 
in which the supersymmetry and the fermionic variables
are represented in a particularly clear manner.
Fermionic and bosonic variables are treated on an equal footing 
in this formalism.

The realisation can be considered as a supersymmetric generalisation 
of the standard definition of the $AdS_5$ and the $S^5$ spaces
(with radii $R$)
by embedding them into two six-dimensional flat spaces,
\begin{equation}
\eta_{\dI \dJ} X^{\dI} X^{\dJ} = -R^2,
\label{RFConstrVecX}
\end{equation}
\begin{equation}
\eta_{I' J'} Y^{I'} Y^{J'} = R^2.
\label{RFConstrVecY}
\end{equation}
Here $X^{\dI}$'s and $Y^{I'}$'s are six-dimensional real vectors.
The indices $\dI, \dJ=0, 1, \ldots, 5$ refer to $SO(4,2)$ vector indices 
and the metric is given by
$\eta_{\dI \dJ}= \diag(-1, 1, 1, 1, 1, -1)$;
$I',J'=1,2,\ldots,6$ are $SO(6)$ vector indices, 
$\eta_{I'J'}=\diag(1, \ldots, 1)$.
The manifolds defined by these equations are equivalent
to the coset spaces $SO(4,2)/SO(4,1)$ and $SO(6)/SO(5)$ respectively
up to global issues which we shall ignore throughout this note.
This type of embedding by quadratic constraints
is often useful, in particular, to make the symmetry properties 
more transparent, as was originally pointed out by Dirac~\cite{RBDirac}.

In our construction we will use linear representations
(more precisely linear ray representations)
of $PSU(2,2|4)$ 
and introduce certain quadratic constraints on the 
representation spaces.
The supermanifold defined by these constraints will 
be shown to be equivalent to $\SuSp$. %(up to two overall $U(1)$ factors).

\paragraph{Representations}
As the supersymmetrisations of $X^{\dI}$ and $Y^{I'}$,
we use two sets of variables $X^{AB}$ and $Y^{AB}$.
They belong to the (super-)anti-symmetric and symmetric products
of the fundamental ray representations of $PSU(2,2|4)$,
\begin{equation}
X^{AB}= -(-1)^{AB} X^{BA},
\end{equation}
\begin{equation}
Y^{AB}= +(-1)^{AB} Y^{BA}.
\end{equation}
Our notation is as follows.
Indices $A, B, \ldots$ are those for the
$PSU(2,2|4)$ fundamental ray representation and take eight values.
They consist of four $SU(2,2)$ ``bosonic'' components $\da, \db, \ldots$
and four $SU(4)$ ``fermionic'' components $a', b', \ldots$.~\footnote{
The assignment of the odd Grassmann parity
to the $SU(4)$ part is purely conventional.
}
The $A, B, \ldots$ indices on the exponent of $(-1)$ should be 
understood as either $0$ for the bosonic components 
or $1$ for 
the fermionic components.
More explicitly we have
\begin{equation}
X^{\da \db}= - X^{\db \da}
,\qquad
X^{\da b'}= - X^{b' \da}
,\qquad
X^{a' b'}= + X^{b' a'},
\end{equation}
\begin{equation}
Y^{\da \db}= + Y^{\db \da}
,\qquad
Y^{\da b'}= + Y^{b' \da}
,\qquad
Y^{a' b'}= - Y^{b' a'}.
\end{equation}
The $\da \db$ and $a' b'$ components
of $X$'s and $Y$'s 
are commuting
and the $\da b'$ and $a' \db$ components are anti-commuting.

We use two irreducible representations of
$PSU(2,2|4)$, rather than one irreducible 
representation. At first sight, it might seem that the use of 
the two variables ($X$'s and $Y$'s) would make the
superspace a direct product of two superspaces.
Actually, the constraints we introduce below intertwine the two variables
so that the final superspace cannot be written as a direct product
of two spaces.
This is consistent with the fact that while the bosonic 
part of the superspace $AdS_5\times S^5$ is written as a direct product,
the full superspace $\SuSp$ is not.

In order to formulate the constraints we introduce 
further conventions and notations on the supersymmetric tensor 
calculus.
We use the 
standard ``left derivative'' convention
for supersymmetric tensor indices, such that
\begin{equation}
v^A w_A= (-1)^{A^2} w_A v^A
\label{RFContractionConvention}
\end{equation}
is a scalar: the indices $A$ should be
contracted in this manner. In this convention, Kronecker's delta
has the index structure
\begin{equation}
\delta_B{}^A.
\end{equation}
By applying complex conjugation to (\ref{RFContractionConvention})
it follows that
\begin{equation}
\overline{
v^A w_A
}
=
\overline{w_A
}
\overline{
v^A 
}
=
\overline{w}_{\bA}
\overline{
v}^{\bA}
\end{equation}
is a scalar. We have introduced indices $\bA, \bB, \ldots$ by defining
$\overline{w_A}=\overline{w}_{\bA}$, $\overline{v^A}=\overline{v}^{\bA}$. 
The $\bA, \bB, \ldots$ indices should be contracted
in the manner indicated in the above formula.
The fundamental ray representation
of $PSU(2,2|4)$ is equipped with a ``hermitian metric''
\begin{equation}
\eta_{A\bB}=\diag(-1,-1,1,1;1,1,1,1)
\end{equation}
and its inverse
\begin{equation}
\eta^{\bA B}=\diag(-1,-1,1,1;1,1,1,1).
\end{equation}
They can be used to lower and raise the indices.

An element of the fundamental ray representation of
the $PSU(2,2|4)$ supergroup transforms as
\begin{equation}
v^A \mapsto v^B U_B{}^A
\end{equation}
where $U_{\da}{}^{\db}$, $U_{a'}{}^{b'}$'s are commuting
and 
$U_{\da}{}^{b'}$, $U_{a'}{}^{\db}$ are anti-commuting.
The variables $X$, $Y$'s transform under $PSU(2,2|4)$ transformations 
by the transformation rule,
\begin{equation}
X^{AB} 
\mapsto
X^{CD}U_C{}^A U_D{}^B (-1)^{D(A+C)},
\label{RFTrfX}
\end{equation}
\begin{equation}
Y^{AB} 
\mapsto
Y^{CD}U_C{}^A U_D{}^B (-1)^{D(A+C)}.
\label{RFTrfY}
\end{equation}
The condition
\begin{equation}
U_{B}{}^A
\eta_{A\bD} 
\overline{ U_C{}^{D} }
\eta^{\bC E}
=
\delta_B{}^E
\label{RFUCond}
\end{equation}
defines the $U(2,2|4)$ supergroup. 
A further constraint
\begin{equation}
\sdet U =1
\label{RFSCond}
\end{equation}
defines the $SU(2,2|4)$ supergroup~\cite{RBSupergroup}.
Finally, by identifying two $U$'s related by an overall $U(1)$ transformation
\begin{equation}
U\sim e^{i \a} U,
\label{RFPCond}
\end{equation}
we obtain the $PSU(2,2|4)$ supergroup.
This identification implies that the fundamental representation should 
be considered as a ray (or projective) representation,
namely elements of the representation space should be identified as follows
\begin{equation}
v^A \sim e^{i \a} v^A.
\end{equation}
As a consequence, the spaces described by the variables  $X$, $Y$ 
also have natural identifications
\begin{equation}
X^{AB} \sim e^{i \a} X^{AB},
\qquad
Y^{AB} \sim e^{i \b} Y^{AB}.
\label{RFIdentification}
\end{equation}
%They should be considered also as belonging to ray representations.
Alternatively, we may also speak about linear representations of $SU(2,2|4)$ or
$U(2,2|4)$, without introducing the identification, though $\FullSG$ is the physically
interesting case.

We denote the complex conjugate of $X^{AB}$ by 
\begin{equation}
\overline{X^{AB}}= \overline{X}^{\bB \bA}.
\end{equation}
We define $X$ with lower indices by
\begin{equation}
X_{AB}=
(-1)^{(B+\bC)A}
\eta_{B\bC} \eta_{A\bD}
\overline{X}^{\bD\bC}
=
(-1)^{(B+\bC)A}
\eta_{B\bC} \eta_{A\bD}
\overline{X^{C D}}
.
\end{equation}
(The sign factor 
$(-1)^{(B+\bC)A}$ above equals $1$ because $\eta$ is diagonal.)
Similarly, we define
\begin{equation}
Y_{AB}=
(-1)^{(B+\bC)A}
\eta_{B\bC} \eta_{A\bD}
\overline{Y}^{\bD\bC}
=
(-1)^{(B+\bC)A}
\eta_{B\bC} \eta_{A\bD}
\overline{Y^{C D}}
.
\end{equation}

\paragraph{Constraints}
On the space described by the variables $X^{AB}$ and $Y^{AB}$, we introduce
the following quadratic constraints, 
\begin{eqnarray}
&&
X^{AC} Y_{CB} =0, \label{RFSuperConstraintCross}
\\
&&X^{AC} X_{CB} - Y^{AC} Y_{CB} =  (-1)^{AB} R^2 \delta_B{}^A. 
\label{RFSuperConstraintDiag}
\end{eqnarray}
The factor $(-1)^{AB}$
in (\ref{RFSuperConstraintDiag})
is necessary to 
make the index structures of the LHS and the RHS
match.

By construction, the LHS and the RHS of the constraints
have the same transformation properties under $PSU(2,2|4)$ transformations,
which can also be verified directly using (\ref{RFTrfX})-(\ref{RFUCond}).
Hence these constraints have invariant meanings
under $PSU(2,2|4)$ transformations.

The constraints are invariant also
under the two overall $U(1)$ transformations
\begin{equation}
X^{AB}\mapsto e^{i\a} X^{AB},
\qquad 
Y^{AB}\mapsto e^{i\b} Y^{AB}.
\label{RFUOneTransformation}
\end{equation}
Hence the constraints (\ref{RFSuperConstraintCross}),
(\ref{RFSuperConstraintDiag}) 
are consistent with the identifications (\ref{RFIdentification}):
the constraints are correctly defined on the ray representations.~\footnote{
Alternatively one may consider the 
constraints to be invariant under $U(2,2|4)$ transformations,
without introducing the identification (\ref{RFIdentification}).}

\paragraph{Equivalence to $\SuSp$}
We will now show that the supermanifold 
defined by the constraints (\ref{RFSuperConstraintCross}),
(\ref{RFSuperConstraintDiag}) is equivalent 
to the coset superspace $\SuSp$.

Any two points on the supermanifold
which are related by a $\FullSG$ transformation are equivalent.
It is therefore natural to start by choosing a representative point
on the manifold and study the manifold in the vicinity of the point.

We first choose a pair of vectors $X_{(0)}^{\dI}$, $Y_{(0)}^{I'}$
satisfying the constraints (\ref{RFConstrVecX}),(\ref{RFConstrVecY}).
The representative point %$X^{AB}_{(0)}$, $Y^{AB}_{(0)}$
is constructed from $X_{(0)}^{\dI}$, $Y_{(0)}^{I'}$
using 
the Clebsch-Gordan coefficients relating 
$SO(4,2)$ and $SU(2,2)$, 
$\Gamma^{\dI \da\db}$, 
and $SO(6)$ and $SU(4)$,
$\Gamma^{I' a'b'}$,
\begin{eqnarray}
&&X_{(0)}^{\da \db} =  X_{(0) \dI} \Gamma^{\dI \da \db}
=\left(X_{(0)} \cdot \Gamma\right)^{\da \db}
,\qquad
X_{(0)}^{\da b'} =  0, \qquad
X_{(0)}^{a' \db} =  0, \qquad
X_{(0)}^{a' b'} = 0, \label{RFRepPtX}
\\
&&
Y_{(0)}^{\da \db} = 0, \qquad
Y_{(0)}^{\da b'} =  0, \qquad
Y_{(0)}^{a' \db} =  0, \qquad
Y_{(0)}^{a' b'} =  Y_{(0) I'} \Gamma^{I' a' b'}
=\left(Y_{(0)} \cdot \Gamma\right)^{a' b'}.
\label{RFRepPtY}
\end{eqnarray}
This point in the superspace satisfies 
the constraints 
(\ref{RFSuperConstraintCross}),(\ref{RFSuperConstraintDiag}),~\footnote{
We fixed the relative sign factors in (\ref{RFSuperConstraintDiag}) by the requirement
that the point specified by 
(\ref{RFRepPtX}),(\ref{RFRepPtY}), (\ref{RFConstrVecX}),(\ref{RFConstrVecY}) is a solution to
(\ref{RFSuperConstraintDiag}).
}
which can be checked using 
\begin{eqnarray}
&&
X_{(0) \da\db}
=
-X_{(0)\dI}
\Gamma^{\dI}{}_{\da\db}
=
-\left(X_{(0)} \cdot \Gamma\right)_{\da \db}
,
\\
&&
Y_{(0) a'b'}
=
Y_{(0)I'}
\Gamma^{I'}{}_{a'b'}
=
\left(Y_{(0)} \cdot \Gamma\right)_{a' b'},
\\
&&
\left(X_{(0)} \cdot \Gamma\right)^{\da \db}
\left(X_{(0)} \cdot \Gamma\right)_{\db \dc}
=
X_{(0)}^{\dI} X_{(0)\dI} \d^{\da}_{\dc}
= -R^2\d^{\da}_{\dc},
\\
&&
\left(Y_{(0)} \cdot \Gamma\right)^{a' b'}
\left(Y_{(0)} \cdot \Gamma\right)_{b' c'}
=
Y_{(0)}^{I'} Y_{(0)I'} \d^{a'}_{c'}
= R^2 \d^{a'}_{c'},
\\
&&
X_{(0)}^{\da \db} X_{(0) \db\dc}= R^2\d^{\da}_{\dc}
,\qquad\quad
Y_{(0)}^{a' b'} Y_{(0) b'c'}= R^2 \d^{a'}_{c'}.
\label{RFXYZeroSquare}
\end{eqnarray}
These formulae follow from properties of the Clebsch-Gordan coefficients summarised in the
appendix.
It is sometimes useful to specify further the point by choosing
$X^{\dI}_{(0)}=(0,\ldots,0,1)$, $Y^{I'}_{(0)}=(0,\ldots,0,1)$.

%%%%%%%%%%%ORBIT
We next consider the orbit of this representative point under all possible 
$\FullSG$ transformations. 
The equivalence of the constrained superspace to
the coset superspace will be shown in two steps.
First we will show the equivalence of the coset space and the orbit space
and then the equivalence of the orbit space and the
constrained superspace.

The orbit and the coset are equivalent if the subgroup of $PSU(2,2|4)$ which leaves 
the representative point fixed is precisely $SO(4,1)\times SO(5)$.

It is sufficient to consider the infinitesimal transformations 
of the representative point specified by $(X_{(0)}^{AB}, Y_{(0)}^{AB})$. An infinitesimal 
transformation $U_A{}^B=\d_A{}^B+\d U_A{}^B$ satisfies,
from $(\ref{RFUCond})$,
\begin{equation}
0=\d U_A{}^C\eta_{C\bB}
+
\eta_{A\bC} 
\overline{\d U_{B}{}^C}.
\label{RFInfinitesimalU}
\end{equation}
The infinitesimal transformation rules of $X$'s and $Y$'s are derived from 
(\ref{RFTrfX}),(\ref{RFTrfY}),
\begin{equation}
\d X^{AB}
=
X^{AC} \d U_C{}^B
+
X^{CB} \d U_C{}^A
(-1)^{(C+A)B},
\end{equation}
\begin{equation}
\d Y^{AB}
=
Y^{AC} \d U_C{}^B
+
Y^{CB} \d U_C{}^A
(-1)^{(C+A)B}.
\end{equation}
The bosonic transformations consist of
$SU(2,2)$ transformations acting on $\da, \db$ indices 
and $SU(4)$ transformations acting on $a', b'$ indices.~\footnote{
The two $U(1)$ transformations in $U(2,2|4)$ which are eliminated
by the $P$ and $S$ conditions (\ref{RFSCond}), (\ref{RFPCond}) 
are absorbed precisely by the identifications (\ref{RFIdentification}) or
the transformations (\ref{RFUOneTransformation}).}
The only non-zero components on which 
a $SU(2,2)$ transformation can act are $X_{(0)}^{\da \db}$.
By the standard property of the Clebsch-Gordan coefficients, 
this action is equivalent to the action of the corresponding $SO(4,2)$
transformation on $X_{(0)}^{\dI}$.
The $SO(4,2)$ transformations which leave $X_{(0)}^{\dI}$ 
(satisfying (\ref{RFConstrVecX})) invariant are precisely 
those forming $SO(4,1)$.
Similarly, the subset of $SU(4)$ transformations 
which leave $X_{(0)}^{AB}$'s and $Y_{(0)}^{AB}$'s invariant are equivalent to
$SO(5)$ transformations which leave $Y_{(0)}^{I'}$ invariant.

It therefore remains to be shown that under any fermionic transformations
(with parameters $\d U_{\da}{}^{b'}$, 
$\d U_{a'}{}^{\db}$, satisfying (\ref{RFInfinitesimalU})), the representative point is not fixed.
The representative point transforms under the fermionic transformations as,
\begin{eqnarray}
&&
\d X^{\da b'}
=
X_{(0)}^{\da\dc}\d U_{\dc}{}^{b'}
=
-\d X^{b'\da},
\label{RFFermionicSUSYVarX}
\\&&
\d Y^{\da b'}
=
-Y_{(0)}^{c'b'}\d U_{c'}{}^{\da}
=
+\d Y^{b'\da}.
\label{RFFermionicSUSYVarY}
\end{eqnarray}
Since $X_{(0)}^{\da \db}$
and 
$Y_{(0)}^{a' b'}$ are invertible (see (\ref{RFXYZeroSquare})),
it follows that the representative 
point is not fixed under any fermionic transformations.
Thus the equivalence between the orbit and the coset is established.

%%%%%%%%%%%%%%%LINEARISATION
From the covariance of the constraints (\ref{RFSuperConstraintCross}),
(\ref{RFSuperConstraintDiag}),
it follows that all points on the orbit space will satisfy the constraints.
Therefore 
the orbit space is contained in the space defined by the constraints.

Hence, in order to show that the orbit space and the constrained manifold 
are equivalent, it is sufficient to check that the constrained manifold
does not contain ``extra directions''.
Hence establishing that the constrained manifold contains the correct
number of bosonic and fermionic dimensions is enough to ensure
the equivalence of the constrained superspace and the orbit space.

Since all points on the manifold will be equivalent,
it suffices to check this property
in the vicinity of the representative point,
\begin{equation}
X^{AB}=X_{(0)}^{AB}+\d X^{AB}
,\qquad
Y^{AB}=Y_{(0)}^{AB}+\d Y^{AB}.
\end{equation}
The constraint (\ref{RFSuperConstraintCross}) can be linearised to yield,
\begin{eqnarray}
&&0= X_{(0)}^{\da \dc} \d Y_{\dc\db}, \label{RFEliminatedYdd}
\\&&
0=
\d X^{\da c'} Y_{(0) c' b'}
+
X_{(0)}^{\da \dc} \d Y_{\dc b'}, \label{RFLinFermiC1}
\\&&
0= \d X^{a' c'} Y_{(0) c'b'}, \label{RFEliminatedXpp}
\end{eqnarray}
and (\ref{RFSuperConstraintDiag}) gives
\begin{eqnarray}
&&
0=
\d X^{\da\dc} X_{(0) \dc\db}
+
X_{(0)}^{\da \dc} \d X_{\dc\db},
\label{RFLinVecX}
\\&&
0=
X_{(0)}^{\da\dc} \d X_{\dc b'}
-
\d Y^{\da c'} Y_{(0) c'b'}, \label{RFLinFermiC2}
\\&&
0
=
\d X^{a' \dc} X_{(0) \dc \db}
-
Y_{(0)}^{ a'c'} \d Y_{c' \db},
\label{RFLinFermiC3}
\\&&
0
=
-
\d Y^{a' c'} Y_{(0) c' b'}
-
Y_{(0)}^{a' c' } \d Y_{c' b'}.
\label{RFLinVecY}
\end{eqnarray}
The formulae (\ref{RFEliminatedYdd}) and (\ref{RFEliminatedXpp}) mean that 
the unwanted components belonging to 
the ten-dimensional symmetric representations of $SU(4)$ (in $\d X$)
and of $SU(2,2)$ (in $\d Y$) are actually eliminated by the constraints.
%since $X_{(0)}^{\da\db}$ and $Y_{(0)}^{a'b'}$ are invertible.
In order to understand the meaning of (\ref{RFLinVecX}), (\ref{RFLinVecY}),
we write
\begin{equation}
\delta X^{\da \db}
=
\delta X^{\dI}
\Gamma_{\dI}{}^{\da\db}
,\qquad
\delta Y^{a' b'}
=
\delta Y^{I'}
\Gamma_{I'}{}^{a'b'},
\end{equation}
using the fact that each of $\Gamma^{\dI \da \db}, \Gamma^{I' a' b'}$ spans
a basis of $4\times 4$ anti-symmetric matrices.
In terms of this notation (\ref{RFLinVecX}), (\ref{RFLinVecY}) imply
\begin{eqnarray}
&&
\d X^{\dI} \Gamma_{\dI}{}^{\da\dc}
X_{(0)}^{\dJ} \Gamma_{\dJ \dc\db}
+
X_{(0)}^{\dI} \Gamma_{\dI}{}^{\da\dc}
\overline{\d X^{\dJ}} \Gamma_{\dJ\dc\db}
=0,
\\&&
\d Y^{I'} \Gamma_{I'}{}^{a'c'}
Y_{(0)}^{J'} \Gamma_{J' c'b'}
+
Y_{(0)}^{I'} \Gamma_{I'}{}^{a'c'}
\overline{\d Y^{J'}} \Gamma_{J'c'b'}
=0.
\end{eqnarray}
Here $\d X^{\dI}$ and $\d Y^{I'}$ are six-dimensional complex vectors.
By decomposing them into real and imaginary parts we obtain,
\begin{eqnarray}
&&
X_{(0) \dI} \Re \d X^{\dI}=0
,\qquad
Y_{(0) I'} \Re \d Y^{I'}=0,
\\&&
X_{(0)}^{\dI} \Im \d X^{\dJ}
-
X_{(0)}^{\dJ} \Im \d X^{\dI}
=0
,\qquad
Y_{(0)}^{I'} \Im \d Y^{J'}
-
Y_{(0)}^{J'} \Im \d Y^{I'}
=0.
\end{eqnarray}
Thus, the imaginary parts of the vectors
$\d X^{\dI}$ and $\d Y^{I'}$ are proportional to
$X_{(0) \dI}$ and $Y_{(0) I'}$ respectively; they are related to 
the original representative point by infinitesimal $U(1)$ transformations
(\ref{RFUOneTransformation}) and therefore should be neglected in the ray representations.
The real parts of the vectors 
$\d X^{\dI}$ and $\d Y^{I'}$ are orthogonal to
$X_{(0) \dI}$ and $Y_{(0) I'}$ respectively; they are nothing but
the tangent spaces of $AdS_5$ and $S^5$ at the representative point.
Thus the bosonic tangent space of the constrained manifold is 
just as it should be.

The constraints for the fermionic components
(\ref{RFLinFermiC1}),
(\ref{RFLinFermiC2}),
(\ref{RFLinFermiC3})
are actually all equivalent 
to
\begin{equation}
\d Y_{\dd b'}= 
-\frac{1}{R^2}
X_{(0) \dd \da}
\d X^{\da c'}
Y_{(0) c' b'}.
\label{RFLinFermiCFin}
\end{equation}
Before imposing the constraints,
the independent fermionic fluctuations
$\d X^{\da b'}$ and $\d Y^{\da b'}$
have 32 complex components.
The above constraint imposes a certain reality condition on them.
Because of this we have 32 real components,
which is the correct number for 
%the type II superstring theory in ten dimensions.
the superspace under consideration.
Hence the constrained supermanifold captures
correctly fermionic directions of the orbit, 
or equivalently the coset space.~\footnote{
One can also check this more directly.
The supersymmetry variation
of the representative point (\ref{RFFermionicSUSYVarX}), (\ref{RFFermionicSUSYVarY}) 
satisfies (\ref{RFLinFermiCFin}) under the condition (\ref{RFInfinitesimalU}).
Conversely, for any variation $\d X$ and $\d Y$ satisfying (\ref{RFLinFermiCFin}),
one can find the fermionic infinitesimal parameters satisfying (\ref{RFInfinitesimalU})
which produce the variation by (\ref{RFFermionicSUSYVarX}), (\ref{RFFermionicSUSYVarY}).
}
Thus finally the equivalence of
$\SuSp$ and the supermanifold defined by the
constraints (\ref{RFSuperConstraintCross}),
(\ref{RFSuperConstraintDiag}) is established.~\footnote{
Alternatively, if one does not introduce the identification (\ref{RFIdentification}),
the same argument presented here establishes the equivalence 
of the constrained space to $\frac{U(2,2|4)}{SO(4,1)\times SO(5)}$.
}

\paragraph{Discussion}
It should be possible to write down the superstring Green-Schwarz action
using the variables $X^{AB}$, $Y^{AB}$ as fields defined on 
the string worldsheet.
The constraints should be imposed by
introducing $\delta$-functionals 
associated with the constraints, in 
the path integral in terms of the fields $X^{AB}$ and $Y^{AB}$.
It may also be possible to take a linear sigma model type approach,
in which one first studies unconstrained $X$, $Y$ fields with 
various coupling constants, and take an appropriate limit of these coupling constants
to realise the constraints.
One should take into account of the $U(1)\times U(1)$ identifications
(\ref{RFIdentification}) in order to ensure that no extra degrees of freedom enter.
%This may be achieved by ``gauging away'' the corresponding degrees of freedom.
It may also be possible to (partially) eliminate the $U(1)$ degrees 
of freedom by introducing non-quadratic constraints constructed 
using the super-determinant.

It is interesting to 
study variables similar to the ones presented in this note
for other AdS superspaces,
in particular those associated with the
supermembrane theory on the $AdS_4\times S^7$ 
and $AdS_7\times S^4$ spaces~\cite{RBDeWitPeetersPlefkaSevrin}.

We hope that the present formulation may
provide a point of view which simplifies 
and clarifies the structure of
supersymmetric theories on AdS spacetimes.
The formalism may also be useful for study 
of quantities controlled by the $PSU(2,2|4)$ symmetry
such as observables in $\mathcal{N}=4$ Super-Yang Mills theory in four-dimension.

This note presents results of work done 
several years ago.
I was stimulated to write up the present results
by two very recent papers \cite{RBSchwarz, RBSiegel} 
which develop a new formulation of 
superstring theory on $AdS_5\times S^5$ using
a parametrisation of the superspace built
along similar directions to the approach proposed in this note. 
The bosonic degrees of freedom in \cite{RBSchwarz, RBSiegel}
are represented in a similar way as 
done in (\ref{RFRepPtX}), (\ref{RFRepPtY}), where we specify 
a part of the bosonic coordinates of the representative point.
The fermionic degrees of freedom are however
introduced differently in our formalism compared to that of \cite{RBSchwarz, RBSiegel}.
The supersymmetry is realised on the (constrained) coordinates of our superspace 
in a linear fashion, whereas in \cite{RBSchwarz, RBSiegel} a non-linear
realisation of the supersymmetry is used.
The formalism presented here may be advantageous for some applications,
as in particular in quantum field theories linearly realised symmetries
can often be more straightforwardly dealt with compared to 
non-linearly realised symmetries.

\paragraph{Acknowledgement}
First of all I would like to thank Stefano Kovacs for invaluable 
discussion, comments, encouragement, and careful reading of the manuscript.
The main part of this work
was done while I was at the Yukawa Institute of Physics
and the Max-Planck Institute for Gravitational Physics. 
I would like to thank colleagues there, in particular, Sergei Frolov, 
Hiroshi Kunitomo, Tristan McLoughlin, Hermann Nicolai,
Jan Plefka, Shigeki Sugimoto, Stefan Theisen, Tatsuya Tokunaga 
for useful discussion, comments and encouragement.
I would like also to thank Yuri Aisaka, Mitsuhiro Kato, 
Yoichi Kazama, Hikaru Kawai, Shota Komatsu,
Tetsuyuki Muramatsu, Yuki Sato, Fumihiko Sugino
for useful discussion, comments and encouragement.
I would like to thank Warren Siegel for interesting comments.
\paragraph{Appendix}
We use
$\Gamma^{\dI \da \db}$,
$\Gamma^{\dI}_{\da \db}$,
for Clebsch-Gordan coefficients relating
$SO(4,2)$ and $SU(2,2)$,
and
$\Gamma^{I' a' b'},
\Gamma^{I'}_{a' b'}
$ for those relating
$SO(6)$ and $SU(4)$. They can be considered as
$4\times 4$ sub-matrices of the $8\times 8$ 
$SO(4,2)$ and $SO(6)$ Gamma matrices.
They are anti-symmetric,
\begin{eqnarray}
&&
\Gamma^{\dI \da \db}
=
-\Gamma^{\dI \db \da},
\qquad
\Gamma^{\dI}{}_{ \da \db}
=
-\Gamma^{\dI}{}_{ \db \da},
\\&&
\Gamma^{I' a' b'}
=
-\Gamma^{I' b' a'},
\qquad
\Gamma^{I'}{}_{ a' b'}
=
-\Gamma^{I'}{}_{ b' a'},
\end{eqnarray}
and satisfy
\begin{eqnarray}
&&
\Gamma^{\dI \da \db}
\Gamma^{\dJ}{}_{\db \dc}
+
\Gamma^{\dJ \da \db}
\Gamma^{\dI}{}_{\db \dc}
=
2 \eta^{\dI\dJ}\delta^{\da}_{\dc}
,\qquad
\Gamma^{I' a' b'}
\Gamma^{J'}{}_{b' c'}
+
\Gamma^{J' a' b'}
\Gamma^{I'}{}_{b' c'}
=
2 \eta^{I'J'}\delta^{a'}_{c'},
\\&&
\eta_{\dc\bar{\da}} 
\eta_{\dd\bar{\db}} 
\overline{\Gamma^{\dI \da \db}}
=
\Gamma^{\dI}{}_{\dc\dd}
,\qquad
\eta_{c'\bar{a'}} 
\eta_{d'\bar{b'}} 
\overline{\Gamma^{I' a' b'}}
=
-
\Gamma^{I'}{}_{c'd'}.
\end{eqnarray}
The matrices
\begin{equation}
\Gamma^{\dI\dJ\da}{}_{\db}
=
\frac{1}{2}
\left(
\Gamma^{\dI\da\dc}
\Gamma^{\dJ}{}_{\dc\db}
-
\Gamma^{\dJ\da\dc}
\Gamma^{\dI}{}_{\dc\db}
\right)
\end{equation}
are linearly independent, and so are
\begin{equation}
\Gamma^{I'J'a'}{}_{b'}
=
\frac{1}{2}
\left(
\Gamma^{I'a'c'}
\Gamma^{J'}{}_{c'b'}
-
\Gamma^{J'a'c'}
\Gamma^{I'}{}_{c'b'}
\right).
\end{equation}
An explicit representation is,
\begin{eqnarray}
&&
\Gamma^{\dI \da \db}
=
\left(
i 1\otimes\s^{2},
-i \s^{1}\otimes\s^{2},
\s^{2}\otimes\s^{1},
\s^{2}\otimes\s^{3},
i \s^{2}\otimes1,
- \s^{3}\otimes\s^{2}
\right),
\\&&
\Gamma^{\dI}{}_{\da \db}
=
\left(
i 1\otimes\s^{2},
i \s^{1}\otimes\s^{2},
\s^{2}\otimes\s^{1},
\s^{2}\otimes\s^{3},
-i \s^{2}\otimes1,
 \s^{3}\otimes\s^{2}
\right),
\\&&
\Gamma^{I' a' b'}
=
\left(
i \s^{2}\otimes1,
\s^{2}\otimes\s^{3},
\s^{2}\otimes\s^{1},
-i \s^{1}\otimes\s^{2},
i \s^{3}\otimes\s^{2},
1\otimes\s^{2}
\right),
\\&&
\Gamma^{I'}{}_{a' b'}
=
\left(
-i \s^{2}\otimes1,
\s^{2}\otimes\s^{3},
\s^{2}\otimes\s^{1},
i \s^{1}\otimes\s^{2},
-i \s^{3}\otimes\s^{2},
1\otimes\s^{2}
\right).
\end{eqnarray}


\begin{thebibliography}{3}
%\cite{Maldacena:1997re}
%\bibitem{Maldacena:1997re}
\bibitem{RBMaldacena}
  J.~M.~Maldacena,
   ``The Large N limit of superconformal field theories and supergravity'',
   {\it Adv. Theor. Math. Phys.}  {\bf 2} (1998) 231
        [hep-th/9711200].
          %%CITATION = HEP-TH/9711200;%%

\bibitem{RBMetsaevTseytlin}
%\cite{Metsaev:1998it}
%\bibitem{Metsaev:1998it}
  R.~R.~Metsaev and A.~A.~Tseytlin,
    ``Type IIB superstring action in AdS(5) x S**5 background,''
      Nucl.\ Phys.\ B {\bf 533} (1998) 109
        [hep-th/9805028].
          %%CITATION = HEP-TH/9805028;%%

\bibitem{RBGreenSchwarz}
%\cite{Green:1983wt}
%\bibitem{Green:1983wt}
  M.~B.~Green and J.~H.~Schwarz,
    ``Covariant Description of Superstrings,''
      Phys.\ Lett.\ B {\bf 136} (1984) 367.
        %%CITATION = PHLTA,B136,367;%%

%\cite{Green:1983sg}
%\bibitem{Green:1983sg}
  M.~B.~Green and J.~H.~Schwarz,
    ``Properties of the Covariant Formulation of Superstring Theories,''
      Nucl.\ Phys.\ B {\bf 243} (1984) 285.
        %%CITATION = NUPHA,B243,285;%%

\bibitem{RBGrisaruHoweMezincescuNilssonTownsend}
%\bibitem{Grisaru:1985fv}
  M.~T.~Grisaru, P.~S.~Howe, L.~Mezincescu, B.~Nilsson and P.~K.~Townsend,
    ``N=2 Superstrings in a Supergravity Background,''
      Phys.\ Lett.\ B {\bf 162} (1985) 116.
        %%CITATION = PHLTA,B162,116;%%


\bibitem{RBDirac}
%\cite{Dirac:1935zz}
%\bibitem{Dirac:1935zz}
  P.~A.~M.~Dirac,
    ``The Electron Wave Equation in De-Sitter Space,''
      Annals Math.\  {\bf 36} (1935) 657.
        %%CITATION = ANMAA,36,657;%%

%\bibitem{Dirac:1936fq}
  P.~A.~M.~Dirac,
    ``Wave equations in conformal space,''
      Annals Math.\  {\bf 37} (1936) 429.
        %%CITATION = ANMAA,37,429;%%

\bibitem{RBSupergroup}
%\cite{Frappat:1996pb}
%\bibitem{Frappat:1996pb}
  See, for example, L.~Frappat, P.~Sorba and A.~Sciarrino,
    ``Dictionary on Lie superalgebras,''
      hep-th/9607161.
        %%CITATION = HEP-TH/9607161;%%
\bibitem{RBDeWitPeetersPlefkaSevrin}
%\cite{deWit:1998yu}
%\bibitem{deWit:1998yu}
  B.~de Wit, K.~Peeters, J.~Plefka and A.~Sevrin,
    ``The M theory two-brane in AdS(4) x S**7 and AdS(7) x S**4,''
      Phys.\ Lett.\ B {\bf 443} (1998) 153
        [hep-th/9808052].
          %%CITATION = HEP-TH/9808052;%%
\bibitem{RBSchwarz}
%\cite{Schwarz:2015lla}
%\bibitem{Schwarz:2015lla}
  J.~H.~Schwarz,
    ``New Formulation of the Type IIB Superstring Action in $AdS_5 \times S^5$,''
      arXiv:1506.07706 [hep-th].
        %%CITATION = ARXIV:1506.07706;%%
\bibitem{RBSiegel}
%\cite{Siegel:2015qka}
%\bibitem{Siegel:2015qka}
  W.~Siegel,
    ``Parametrization of cosets for AdS5xS5 superstring action,''
      arXiv:1506.08172 [hep-th].
  %%CITATION = ARXIV:1506.08172;%%
\end{thebibliography}
\end{document}